# A Survey of Security Assessment Ontologies

Ferrucio de Franco Rosa[1,2], Mario Jino[2]

[1] Renato Archer Information Technology Center (CTI),
Campinas-SP, Brazil
[2] School of Electrical and Computer Engineering at University of Campinas (UNICAMP),
Campinas-SP, Brazil
ferrucio.rosa@cti.gov.br, jino@dca.fee.unicamp.br

**Abstract.** A literature survey on ontologies concerning the Security Assessment domain has been carried out to uncover initiatives that aim at formalizing concepts from the "Security Assessment" field of research. A preliminary analysis and a discussion on the selected works are presented. Our main contribution is an updated literature review, describing key characteristics, results, research issues, and application domains of the papers. We have also detected gaps in the Security Assessment literature that could be the subject of further studies in the field. This work is meant to be useful for security researchers who wish to adopt a formal approach in their methods.

## 1  Introduction

Security defects, whose corrections are already known, continue to be introduced in new systems or exist in systems in operation, just waiting to be activated. On the other hand, it is not easy to detect and fix unknown security problems systematically. Activities to test the security of information systems often rely heavily on knowledge and experience of professionals involved in these activities [1–7].

Main research directions in Security Assessment and efforts aimed at formalizing the conceptual domain should be uncovered. A literature review is the tool to enable us to identify research issues, contributions, characteristics and objectives of these efforts. Analysis of the selected works is as important as the literature review itself; the foundational concepts of the domain and their relationships must be uncovered. Domain knowledge is necessary to formalize research in a systematic way.

The knowledge available in the literature supporting security assessment is not structured enough. Approaches based on a conceptual formalization, properly built into the information security context, can provide a better support for security assessment.

Ontologies can be used as a vocabulary, a dictionary or a roadmap of the Information Security domain. Furthermore, an ontology can be used to reason (provide inferences) about relationships between entities [8].

Hence, our main objective is to search for initiatives aimed at formalizing concepts from the "Security Assessment" fields of research, by means of ontologies.

## 2 Survey

The literature review process is based, with adaptations, on the guidelines for performing systematic reviews proposed by Biolchini [9] and Kitchenham [10].

We present the literature review on "Information Systems Security Assessment" aimed to be used as a supporting resource by researchers in the field. Specifically, we are looking for papers in the literature representing the state of the art in two search fronts. The first front aims to identify papers that systematize and formalize concepts on "Information Security", by means of ontologies. In the second front, we include "Test and Assessment" and the corresponding related keywords.

We applied a systematic approach in the search process by: selecting search databases, choosing keywords, defining search strings, and specifying inclusion and exclusion criteria. 80 papers were identified and 47 were selected.

Seven databases were selected for this study: IEEE Xplore, ACM Digital Library, Scielo, Proquest, ScienceDirect – Elsevier, SpringerLink, and Google Scholar. In the Search Databases (advanced search), the following parameters were considered: "title", "abstract", "entire document" and their combinations, if available in the search database.

Concerning selection criteria, we used: (i) Inclusion criteria: Recent works; Works containing more citations; Works containing important concepts; Works related to the defined research questions. (ii) Exclusion criteria: Works not related to the defined research questions.

We have defined keywords and a search string. English is used in the search. Table 1 presents the keywords and the search string.

**Table 1.** Keywords and Search String.

| Keyword | Search String |
| --- | --- |
| Security; Privacy; Dependability; Ontology; Test; Testing; Assessment; Criterion; Analysis; Audit; Evaluation. | ((Security OR Privacy OR Dependability) AND (Ontology) AND (Test OR Testing OR Assessment OR Analysis OR Audit OR Evaluation)) |

## 3  Research Works on Ontologies

Most of the identified works aim to describe the domains of Software Security and Software Test, including their various sub-domains (e.g.: Risk Management, Security Policies; Incident Analysis; Attack Patterns, Performance Tests; Expert Systems Tests, etc.). Works which aim to describe domains and sub-domains are the following: [11]; [12]; [13]; [14]; [15, 16]; [17]; [18]; [19]; [20]; [21]; [22]; [23]; [24]; [25]; [26]; [27]; [28]; [29]; [30]; [31]; [32]; [33]; [34]; [35]; [36]; [37]; [38]; [39]; [40]; [41]; [42]; [43]; [44]; [45]; [44]; [8]; [46]; [47].

Generic and abstract proposals (Top-Level Ontologies) can be found in [48], [3], [49], [50], and [51]. Specific proposals (Task and Application Ontologies) can be found in [52], [53], and [54].

After analysis of the selected works, the following can be highlighted:

In [47] Raskjn et al. present concepts of the Information Security domain, and also explains how ontologies can be used to support the Information Security field, in order to provide a theoretical basis. In [54] Viljanen presents an ontology of Trust, in order to facilitate interoperability between systems. A common vocabulary is proposed to describe facts that should be considered in trust calculation.

In [8] an information security ontology in OWL is presented by Herzog et al.. This work is aimed at modeling the main concepts of the domain. The authors describe content, manner of use, possibility of extension, technical implementation and tools to handle the ontology. In [41] Fenz & Ekelhart present an ontology of the Information Security domain, focused on Risk Management. The authors use the German IT Grundschutz Manual [55] and the NIST Handbook [56] as references. The ontology uses the concepts of threat, vulnerability and control to represent knowledge in the information security domain.

In [33] Evesti et al. present an ontology to support the process of measuring Information Security, whereas Feledi & Fenz [22] present a formalization of information security knowledge, by means of an ontology. According to the authors, we need to make explicit knowledge, so that it can be incorporated and used by both humans (human-readable format) and machines (machine-readable format).

A top-level ontology of security requirements is presented in [3] by Salini & Kanmani. Based on this ontology, we can design and develop requirements for electronic voting systems (e-voting). The main objective of this work is to propose security patterns to facilitate the process of identifying security requirements for e-voting systems. The authors present specific security properties for e-voting systems, namely: anonymity, disclosability, uniqueness, accuracy, transparency, and non-coercibility.

In [19] Gyrard et al. present the STACK ontology (Security Toolbox: Attacks & Countermeasures) to aid developers in the design of secure applications. STACK defines security concepts such as attacks, countermeasures, security properties and their relationships. Countermeasures can be cryptographic concepts (encryption algorithm, key management, digital signature, and hash function), security tools, or security protocols. In [18] Kotenko et al. present an ontology of security metrics, specifically built for the SIEM (Security Information and Event Management) domain.

A standard-based security ontology is presented by Ramanauskaite et al. in [17]. The authors propose an ontology aimed to cover a larger number of standards. The authors mapped papers Herzog et al. [8] and Fenz, Pruckner, & Manutscheri [57] with the standards ISO 27001 [58], PCI DSS [59], ISSA 5173 [60] and NISTIR 7621 [61].

In [15] Salini & Kanmani present an ontology of security requirements for web applications. This work aims at enabling the reuse of knowledge about security requirements in the development of different web applications. In [52] Khairkar et al. present an ontology to detect attacks on Web systems. The authors use semantic web concepts and ontologies to analyze security logs to identify potential security issues. This work aims to extract semantic relationships between attacks and intrusions in an Intrusion Detection System (IDS).

In [13] Kang & Liang present a security ontology, for use in the software development process. The proposed ontology can be used for identifying security requirements, as a practical and theoretical basis. In [11] Koinig et al. present a security ontology for cloud computing and a brief literature review. The authors consider the regulatory requirements contained in standards such as HIPAA (Health Insurance Portability and Accountability Act) [62], SOX (Sarbanes Oxley Law) [63], and ISO/IEC 27001 [58].

## 4 Discussion and Related Work

Most of the works address the conceptual formalization issue that was identified in 100% of the papers on ontologies up to 2008. After 2008, we can see other applications for ontologies.

Concerning the ontologies, we can identify that among recent works (2011-2015) most address the conceptual formalization issue. However, we can identify works that address other research issues, namely: Requirements [16], [3], [27], [29]; Audit [17], [21], [23], [24]; Knowledge Management [12], [22], [34], [40].

Concerning Application Domain, there is a good distribution among the domains in recent works. Most of the least recent works (up to 2009) are: Security in General View and Knowledge Management. On the other hand, Legal regulation, Detection and Prevention of Intrusions, Embedded Systems and IoT and Mobile Applications are not found in recent works (from 2011 on).

Regarding related work, Blanco et al. [64] present a literature review and proposes a method for integrating ontologies, through qualitative analysis of more mature proposals. In [65] Souag et al. present an analysis of existing security ontologies and their use in defining requirements. The work is part of a project that aims to improve the definition of security requirements using ontologies. This study addresses the question: which security ontology is suitable for my needs? The literature review was adapted from Barnes [66] and Rainer & Miller [67]. Blanco et al. [64] and Souag et al. [65] emphasize the importance of previous literature reviews and point to the need of updates.

## 5 Conclusion

We identified gaps in the Security Assessment literature. There is a lack of works that address the following research issues: Reusing Knowledge; Automating Processes; Increasing Coverage of Assessment; Secure Sharing of Information; Defining Security Standards; Identifying Vulnerabilities; Measuring Security; Protecting Assets; Assessing, Verifying or Testing the Security. This finding indicates that these research issues can be addressed in future papers, increasing the likelihood of original contributions.

Most of works on security ontologies aims to describe the Information Security domain (more generic), or other specific subdomain of security, but not specifically the Security Assessment domain. We identified a lack of ontologies that consider the relation of "Information Security" and "Software Assessment" fields of research. For example, we identified in some analyzed ontologies that there are certain confusion about key concepts of the Security Assessment domain, such as vulnerability, exploit, risk, weakness, verification, certification.

The literature review shows key characteristics, research results, research issues, and application fields of the works. This work is meant to be useful for security researchers who wish to formalize knowledge in their methods.

We are currently working on expanding this survey, to include other main contributions, such as taxonomies, methods, approaches, systems, among other contributions to be identified and categorized.

## References

1. Barros, C.P. de, Rosa, F. de F., Balcão Filho, A.F.: Software Testing With Emphasis on Finding Security Defects. In: IADIS - The 12th International Conference on WWW/Internet. pp. 226–228 (2013).
2. Tsoumas, B., Gritzalis, D.: Towards an ontology-based security management. Proc. - Int. Conf. Adv. Inf. Netw. Appl. AINA. 1, 985–990 (2006).
3. Salini, P., Kanmani, S.: A Knowledge-Oriented Approach to Security Requirements Engineering for E-Voting System. 49, 21–25 (2012).
4. Gartner, S., Ruhroth, T., Burger, J., Schneider, K., Jurjens, J.: Maintaining requirements for long-living software systems by incorporating security knowledge. 2014 IEEE 22nd Int. Requir. Eng. Conf. 103–112 (2014).
5. The MITRE Corporation: Common Vulnerabilities and Exposures (CVE). (2015).
6. Wita, R., Jiamnapanon, N., Teng-amnuay, Y.: An ontology for vulnerability lifecycle. 3rd Int. Symp. Intell. Inf. Technol. Secur. Informatics, IITSI 2010. 553–557 (2010).
7. NIST - US National Institute of Standards and Technology: NVD CVSS - Common Vulnerability Scoring System Support v2. (2015).
8. Herzog, A., Shahmehri, N., Duma, C.: An ontology of information security. Int. J. Inf. Secur. Priv. 1, 1–23 (2007).
9. Biolchini, J., Mian, P.G., Candida, A., Natali, C.: Systematic Review in Software Engineering. Engineering. 679, 165–176 (2005).
10. Kitchenham, B.: Procedures for performing systematic reviews. Keele, UK, Keele Univ. 33, 28 (2004).
11. Koinig, U., Tjoa, S., Ryoo, J.: Contrology - An Ontology-Based Cloud Assurance Approach. 2015 IEEE 24th Int. Conf. Enabling Technol. Infrastruct. Collab. Enterp. 105–107 (2015).
12. Souza, E.F. de, de Souza, E.F.: Knowledge Management Applied to Software Testing: An Ontology Based, (2014).
13. Kang, W., Liang, Y.: A security ontology with MDA for software development. Proc. - 2013 Int. Conf.

Cyber-Enabled Distrib. Comput. Knowl. Discov. CyberC 2013. 67–74 (2013).
14. Freitas, A.L.S. da C.: Ontologia para teste de desempenho de software, https://performance-ontology.googlecode.com/files/Dissertacao_ArturFreitas.pdf, (2013).
15. Salini, P., Kanmani, S.: Ontology-based representation of reusable security requirements for developing secure web applications. Presented at the (2013).
16. Panchal, J., Chirchi, V.R.: Privacy Preservation Requirement for Personal Health Record : A Survey on Security Prototypes. Int. J. Adv. Comput. Eng. Appl. 2, 13–18 (2013).
17. Ramanauskaite, S., Olifer, D., Goranin, N., Čenys, A.: Security ontology for adaptive mapping of security standards. Int. J. Comput. Commun. Control. 8, 878–890 (2013).
18. Kotenko, I., Polubelova, O., Saenko, I., Doynikova, E.: The ontology of metrics for security evaluation and decision support in SIEM systems. Proc. - 2013 Int. Conf. Availability, Reliab. Secur. ARES 2013. 638–645 (2013).
19. Gyrard, A., Bonnet, C., Boudaoud, K., Gyrard, A., Bonnet, C., Boudaoud, K., Stac, T., Toolbox, S., Gyrard, A., Bonnet, C.: The STAC ( Security Toolbox : Attacks & Countermeasures ) ontology. (2014).
20. Bhaumik, A.: An approach in defining Information Assurance Patterns based on security ontology and meta-modeling, (2012).
21. D'Agostini, S., Di Giacomo, V., Pandolfo, C., Presenza, D.: An ontology for run-time verification of security certificates for SOA. Proc. - 2012 7th Int. Conf. Availability, Reliab. Secur. ARES 2012. 525–533 (2012).
22. Feledi, D., Fenz, S.: Challenges of web-based information security knowledge sharing. 2012 Seventh Int. Conf. Availability, Reliab. Secur. 514–521 (2012).
23. Birkholz, H., Sieverdingbeck, I., Sohr, K., Bormann, C.: IO: An interconnected asset ontology in support of risk management processes. Proc. - 2012 7th Int. Conf. Availability, Reliab. Secur. ARES 2012. 534–541 (2012).
24. Diéguez, M., Sepúlveda, S., Cares, C.: On optimizing the path to information security compliance. Proc. - 2012 8th Int. Conf. Qual. Inf. Commun. Technol. QUATIC 2012. 182–185 (2012).
25. Talib, A.M., Atan, R., Abdullah, R., Azrafi, M., Murad, A.: Security Ontology Driven Multi Agent System Architecture for Cloud Data Storage Security : Ontology Development. Int. J. Comput. Sci. Netw. Secur. 12, 63–72 (2012).
26. Nabil, S., Mohamed, B.: Security ontology for semantic SCADA. CEUR Workshop Proc. 867, 179–192 (2012).
27. Lotz, V., Kaluvuri, S.P., Di Cerbo, F., Sabetta, A.: Towards security certification schemas for the internet of services. 2012 5th Int. Conf. New Technol. Mobil. Secur. - Proc. NTMS 2012 Conf. Work. (2012).
28. Massacci, F., Mylopoulos, J., Paci, F., Yu, Y., Tun, T.T.: An Extended Ontology for Security Requirements. Presented at the (2011).
29. Bialas, A.: Common criteria related security design patterns for intelligent sensors-knowledge engineering-based implementation. Sensors. 11, 8085–8114 (2011).
30. Janpitak, N., Sathitwiriyawong, C.: Data Center Physical Security Ontology for Automated Evaluation. Weblidi.Info.Unlp.Edu.Ar. (2011).
31. Nascimento, C., Ferraz, F., Assad, R.: OntoLog: Using Web Semantic and Ontology for Security Log Analysis. Proc. Sixth Int. Conf. Softw. Eng. Adv. ICSEA 2011. 177–182 (2011).
32. Ciuciu, I., Claerhout, B., Schilders, L., Meersman, R.: Ontology-Based Matching of Security Attributes for Personal Data Access in e-Health. Lncs Springer-Verlag Berlin Heidelb. 7045, 605–616 (2011).
33. Evesti, A., Savola, R., Ovaska, E., Kuusijarvi, J.: The design, instantiation, and usage of information security measuring ontology. Proc. 4th IEEE Int. Conf. Self-Adaptive Self-Organizing Syst. 204–212 (2011).
34. Da Silva, P.F., Otte, H., Todesco, J.L., Gauthier, F. a O.: Uma ontologia para gestão de segurança da informaç ão. CEUR Workshop Proc. 776, 141–146 (2011).
35. Basile, C., Silvestro, J., Lioy, A., Canavese, D.: Security Ontology Definition. (2011).
36. Blackwell, C.: A security ontology for incident analysis. Proc. Sixth Annu. Work. Cyber Secur. Inf. Intell. Res. - CSIIRW '10. 1 (2010).
37. Vorobiev, A., Bekmamedova, N.: An ontology-driven approach applied to information security. J. Res. Pract. Inf. Technol. 42, 61–76 (2010).
38. Takahashi, T., Kadobayashi, Y., Fujiwara, H.: Ontological approach toward cybersecurity in cloud computing. Proc. 3rd Int. Conf. Secur. Inf. networks - SIN '10. 100 (2010).
39. Singhal, A., Wijesekera, D.: Ontologies for modeling enterprise level security metrics. Proc. Sixth Annu. Work. Cyber Secur. Inf. Intell. Res. - CSIIRW '10. 1 (2010).
40. Almeida, M.B., Souza, R.R., Coelho, K.C.: Uma Proposta de Ontologia para o Domínio Segurança da

Informação em Organizações: descrição do estágio terminológico. Inf. Soc.Est. 20, 155–168 (2010).
41. Fenz, S., Ekelhart, A.: Formalizing information security knowledge. ... 4th Int. Symp. Inf. .... 183 (2009).
42. Bialas, A.: Ontology-based security problem definition and solution for the common criteria compliant development process. Proc. 2009 4th Int. Conf. Dependability Comput. Syst. DepCos-RELCOMEX 2009. 3–10 (2009).
43. Bezerra, D., Costa, A., Okada, K.: SwTOI (Software Test Ontology Integrated) and its application in Linux test. CEUR Workshop Proc. 460, 25–36 (2009).
44. Azevedo, R.R. De, Almeida, S.C. De, Brasil, P.A., Almeida, M.J.S.C., Filho, E.C.D.B.C.: CoreSec : An Ontology of Security Aplied to the Business Process of Management. (2008).
45. Azevedo, R.R. De: CoreSec : Uma Ontologia para o Domínio de Segurança da Informação, (2008).
46. Barbosa, E.F., Nakagawa, E.Y., Maldonado, J.C.: Towards the Establishment of an Ontology of Software Testing. Seke. (2006).
47. Raskjn, V., Hempelmann, C.F., Nirenburg, S., Lafayette, W.: Ontology in Information Security : A Useful Theoretical Foundation and Methodological Tool. Work. New Secur. Paradig. 53–59 (2002).
48. Souag, A., Salinesi, C., Mazo, R., Comyn-Wattiau, I.: A Security Ontology for Security Requirements Elicitation. Presented at the (2015).
49. Grobler, M., van Vuuren, J.J., Leenen, L.: Implementation of a Cyber Security Policy in South Africa : Reflection on Progress and the Way Forward. ICT Crit. Infrastructures Soc. 386, 215–225 (2012).
50. Zhu, H., Huo, Q.: Developing a software testing ontology in UML for a software growth environment of web-based applications. Softw. Evol. with UML. 1–34 (2005).
51. Jutla, D., Xu, L.: Privacy Agents and Ontology for the Semantic Web. Am. Conf. Inf. Syst. 1760–1767 (2004).
52. Khairkar, A.D., Kshirsagar, D.D., Kumar, S.: Ontology for detection of web attacks. Proc. - 2013 Int. Conf. Commun. Syst. Netw. Technol. CSNT 2013. 612–615 (2013).
53. Liu, F.-H., Lee, W.-T.: Constructing Enterprise Information Network Security Risk Management Mechanism by Ontology. J. Appl. Sci. Eng. 13, 79–87 (2010).
54. Viljanen, L.: Towards an Ontology of Trust. Computer (Long. Beach. Calif). 3592, 175–184 (2005).
55. British Standards Institution (BSI): BSI Standard 100-2 IT-Grundschutz Methodology Version 2.0, (2008).
56. Bowen, P., Hash, J., Wilson, M.: NIST - Information Security Handbook: A Guide for Managers, http://csrc.nist.gov/publications/nistpubs/800-100/SP800-100-Mar07-2007.pdf, (2006).
57. Fenz, S., Pruckner, T., Manutscheri, A.: Ontological mapping of information security best-practice guidelines. Lect. Notes Bus. Inf. Process. 21 LNBIP, 49–60 (2009).
58. ISO/IEC: ISO/IEC 27001:2013 Information technology -- Security techniques -- Information security management systems -- Requirements, (2013).
59. PCI Security Standards Council: Payment Card Industry Data Security Standard (PCI DSS), https://www.pcisecuritystandards.org/.
60. ISSA-UK: ISSA 5173 – The Security Standard for SMES, https://www.2-sec.com/2011/03/22/issa-5173-the-security-standard-for-smes/.
61. NIST - US National Institute of Standards and Technology: NISTIR 7621 - Small Business Information Security: The Fundamentals., http://www.nist.gov.
62. U.S. Department of Health & Human Services: Health Insurance Portability and Accountability Act (HIPAA), http://www.hhs.gov/ocr/privacy/.
63. Addison-Hewitt Associates: A Guide To The Sarbanes-Oxley Act (SOX), http://www.soxlaw.com/.
64. Blanco, C., Lasheras, J., Fernández-Medina, E., Valencia-García, R., Toval, A.: Basis for an integrated security ontology according to a systematic review of existing proposals. Comput. Stand. Interfaces. 33, 372–388 (2011).
65. Souag, A., Salinesi, C., Comyn-Wattiau, I.: Ontologies for security requirements: A literature survey and classification. Lect. Notes Bus. Inf. Process. 112 LNBIP, 61–69 (2012).
66. Barnes, S.J.: Assessing the value of IS journals. Commun. ACM. (2005).
67. Rainer, R.K., Miller, M.D.: Examining differences across journal rankings. Commun. ACM. 48, 91–94 (2005).